\documentclass[aps,pra,twocolumn,showkeys,groupedaddress,nofootinbib]{revtex4-1}

\usepackage{amsmath,amsfonts}
\usepackage{epsfig}

\begin{document}  
\title {\bf An interpretation and understanding of complex modular values} 

\author{ Le Bin Ho}
\thanks{Electronic address: binho@qi.mp.es.oaska-u.ac.jp}
\affiliation{Graduate School of Engineering Science, Osaka University, Toyonaka, Osaka 560-8531, Japan}

\author{ Nobuyuki Imoto}
\affiliation{Graduate School of Engineering Science, Osaka University, Toyonaka, Osaka 560-8531, Japan}

\date{\today}

\begin{abstract}
In contrast to that a weak value of an observable is usually divided into real and imaginary parts, here we show that separation into modulus and argument is important for modular values. We first show that modular values are expressed by the average of dynamic phase factors with complex conditional probabilities. We then relate, using the polar decomposition, the modulus of the modular value to the relative change in the qubit pointer post-selection probabilities, and relate the argument of the modular value to the summation of a geometric phase and an intrinsic phase.
\end{abstract}

\maketitle

{\section{Introduction}\label{sec_int}}
Quantum modular value is a concept that was proposed by Kedem and Vaidman in 2010 \cite{Kedem105}.  When a quantum system is prepared in an initial state $|\psi\rangle$ and post-selected in a final state $|\phi\rangle$, the modular value $(\hat{A})_{\rm m}$ of an observable $\hat{A}$ is defined to be the expectation value of the {\it dynamically evolved phase factor} $e^{-ig\hat A}$, and is expressed as
\begin{align}{\label{modular}}
(\hat{A})_{\rm m} = \dfrac{\langle\phi|e^{-ig\hat A}|\psi\rangle}{\langle\phi|\psi\rangle}\;,
\end{align}
where $g$ is the magnitude of the coupling. 

Eq. (\ref{modular}) is derived, in relation to {\it non}-weak measurement, as follows \cite {Kedem105,Ho}. In between the preparation of $|\psi\rangle$ and the post-selection of $|\phi\rangle$ of the quantum system, a qubit pointer, which is prepared in state $\gamma |0\rangle + \bar\gamma |1\rangle$, is coupled to the system. Here we use the word ``pointer" instead of ``meter" since we use the subscript (and sometimes superscript) m to denote ``modular value", and then, we use p to denote ``qubit pointer" (=meter). Following the standard  von Neumann treatment \cite{vNm}, the interaction Hamiltonian is given by $ \hat H = g(t)\hat A\otimes\hat P$, where $g(t)$ is an arbitrary (possibly time dependence) coupling constant and $\hat P = |1\rangle\langle 1|$ is a projection operator of the qubit pointer. The final state of the qubit pointer after the interaction and post-selection on the system is calculated to be $ \langle \phi|\psi\rangle (\gamma |0\rangle +\bar\gamma(\hat{A})_{\rm m}|1\rangle)$, where  $(\hat{A})_{\rm m} \equiv  \langle\phi|\hat e^{-ig\hat A}|\psi\rangle/\langle\phi|\psi\rangle$, and is named ``modular value". Here, $g$ is the magnitude of the coupling defined as $g \equiv \int g(t) {\rm d}t$.

\sloppy
Apparently, a modular value of an observable is related to the corresponding weak value through $\langle \hat{A}\rangle_{\rm w} = i \left[ {\partial \over \partial g} (\hat{A})_{\rm m} \right]_{g = 0}$, where the weak value, denoted as $\langle \hat{A}\rangle_{\rm w}$, is defined to be the expectation value of the measured values of $\hat{A}$ through repeated weak measurements performed in-between the preparation and the post-selection, which was shown to be $\langle \hat{A}\rangle_{\rm w} = \langle\phi|\hat A|\psi\rangle / \langle\phi|\psi\rangle$ \cite{Aharonov60}. Interestingly, the modular value can be related to the weak value even for nonzero $g$'s \cite{Kedem105, Ho}. 

Weak values can take values outside the range of eigenvalues of $\hat{A}$, and can even be complex. Although this strange property has been discussed for weak values in the context of probabilistic interpretation \cite{Hosoya43} or contextuality \cite{Pusey113}, and is considered to be  a suitable index to describe many intriguing quantum phenomena including quantum paradoxes \cite{Aharonov17,Hardy68,Aharonov301,Lundeen102,Yokota11,Aharonov15,Hasegawa5}, and even is  applied to amplification and precision metrology \cite{Pang113,Zhang114,Huang17,Nishizawa92,Susa92}, discussions on this kind of property are still missing in modular values. Furthermore, as also claimed by us in Ref. \cite{Ho}, modular values are sometimes more beneficial than weak values, in the sense that measuring a modular value is more efficient than measuring a weak value because the measurement coupling constant $g$ can be made large. From the experimental point of view, modular values are seemingly easier to measure because one can simply perform the tomography using binary outcomes in the qubit pointer \cite{Ho}. Therefore, it is necessary and meaningful to shed light on the study of quantum modular values. Specifically, we focus our study on the behavior of quantum modular values as complex numbers in relation with complex conditional probabilities in the present work.

Recently, Cormann et al. have proposed a new procedure to measure a quantum modular value \cite{Cormann}. They demonstrated directly the modulus and argument of the modular value as functions of the measurement strength. These observations require more detailed theory on the behavior of quantum modular values as complex numbers.

The main purpose of this work is to understand the complex behavior of quantum modular values. In contrast to the weak-value case, where the real and imaginary parts play a significant role each, what is more important is the modulus and argument in the modular-value case, as we show in this paper. To see this, we first relate the modular value to complex conditional probabilities as will be seen in Sec. \ref{sec_sta}. Then we derive the polar decomposition (modulus and argument) of the modular value, and relate them to some interesting concepts in Secs. \ref{sec_mod} and  \ref{sec_arg} as below.

We first derive that a modular value can be expressed as the average of a dynamic phase factor over all eigenvalues with complex conditional probabilities [Eq. \eqref{mv}].  Thus, the modular value is interpreted in the context of the complex conditional probabilities, which is analogous to the weak-value case [Eq. \eqref{Aw=sum_a a Pr(a|psi,phi)}]. In addition, the chain rule of modular values Eq. \eqref{modular_rule}, analogous to that for weak values Eq. \eqref{weak_rule}, is also derived. This implies that a modular value can be expressed by a set of intermediate weak values and modular values. These results will be described in Sec. \ref{sec_sta}. 

In Secs. \ref{sec_mod} and  \ref{sec_arg}, we discuss the interpretation of the modulus and argument components of a modular value in the polar decomposition \cite{Botero44,BoteroArxiv}. Therein, in Sec. \ref{sec_mod}, we express the modulus of the modular value in relation to the relative change in the qubit pointer post-selection probabilities. We define {\it the relative change in the qubit pointer post-selection probabilities} (hereafter referred to as ``relative change" for short) as the ratio of the statistical frequency of finding the chosen final state $|1\rangle$ to that of $|0\rangle$ for the qubit pointer.
We show that the modulus of the modular value is proportional to the square root of the corresponding relative change. Using this theory, a qubit system is extensively examined, and  we also propose a way to experimentally determine the coupling constant $g$ (when it is unknown beforehand) by measuring this relative change.

In Sec. \ref{sec_arg}, we relate the argument of the modular value to the Pancharatnam relative phases \cite{Pancharatnam44} by considering the initial state, the evolved state, and the post-selected state of the system. In this work, we pay attention to two kinds of phase, that are, {\it the intrinsic phase} (the phase shift by the evolution of the state) and {\it the geometric phase} (the geometric phase associated with the three states). It is particularly shown that the argument of the modular value is expressed by the total Pancharatnam relative phases (the summation of the geometric phase and the intrinsic phase). 
The Pancharatnam phase plays an important role for robust quantum gates in quantum information processing \cite{Sjoqvist}, and has applications to quantum information such as fault-tolerant quantum computation \cite{Zanardi264, Duan292} and weak measurement \cite{Tamate11}. So the study on modular values in connection with the Pancharatnam phases might open new possibilities for further studies.

Finally, we summarize the results of this paper in Sec. \ref{sec_con}.

{\section{Probabilistic interpretation of modular values}\label{sec_sta}}
In this section, we extend the previous studies of Hofmann about weak values in Refs. \cite{Hofmann13, Hofmann14} to probabilistic interpretation for modular values. We show that, analogous to weak values, modular values can be understood in the context of complex conditional probabilities.  

Let us first give a brief summary to the previous studies. The ordinary expectation value, that is, the expectation value without the post-selection condition, can be interpreted as the average of weak values over all possible post-selection states as $\langle \hat{A}\rangle_{\psi} = \sum_\phi \langle \hat{A}\rangle_{\rm w} \text{Pr}(\phi|\psi)$, where $\text{Pr}(\phi | \psi)$ is the conditional probability of observing state $| \phi \rangle$ on condition that the prepared state is $| \psi\rangle$, and is, of course, equal to $| \langle \phi | \psi \rangle |^2$ \cite{Aharonov72,Tollaksen40,Shikano43,Hosoya43,Hosoya44}. Comparing this to $\bar x = \sum_x x\text{ Pr}(x)$ in standard statistics, the weak value can be treated as a 
$|\phi\rangle$ dependent variable --- which can take complex values, but its expectation value is real with the real probability $\text{Pr}(\phi | \psi)$ \cite{Hosoya43}. 

Interestingly, a weak value itself can be regarded as the average of conditional probabilities --- in this case, however, conditional probabilities themselves can take complex values \cite{Shikano43,Mitchison76,Hofmann13,Hofmann14}. In fact, using the spectral decomposition $\hat A = \sum_a a \ \hat{\Pi}_a$  $\left( \hat{\Pi}_a \equiv |a\rangle\langle a| \right)$, it is straightforward to obtain \cite{Hofmann13,Hofmann14}
\begin{align}{\label{Aw=sum_a a Pr(a|psi,phi)}}
\langle A\rangle_{\rm w} = \sum_a a \ \text {Pr} (a | \psi,\phi)\;. 
\end{align}
Here,
\begin{align}{\label{complex_conditional_prob_1}}
\text{Pr} (a | \psi,\phi) = {\langle\phi | \hat{\Pi}_a | \psi\rangle \over \langle\phi|\psi\rangle} = \langle \hat{\Pi}_a \rangle_{\rm w} \;,
\end{align}
is known as the {\it complex conditional probability} \cite{Hofmann13,Hofmann14}, for the process: from the initially prepared state $|\psi\rangle$ to the finally post-selected state $|\phi\rangle$ via the intermediate state $|a\rangle$. Normally, the weak value of the projection operator $\hat A = |a\rangle\langle a|$ is the transition amplitude from the initial state to the final state via the intermediate state, the squared of its values which is known as the probability \cite{Hosoya43}. However, here we interpreted it as complex conditional probability in the scene that the state $|a\rangle$ is might not observed by projective measurements, for example  ``counter-factual probabilities" \cite{Dirac,Feynman,Hofmann81}.

Using this, the ordinary expectation value is expressed by the chain rule as 
$\langle A\rangle_{\psi} = \sum_a a\sum_\phi \text {Pr} (a | \psi,\phi) \text{Pr}(\phi|\psi) $.
Here, $\text{Pr}(\phi|\psi)$ is real but $\text {Pr} (a | \psi,\phi)$, so to say a jointly conditioned probability, can be negative or even complex, and the finally obtained expectation value is real. Nevertheless, Hofmann also constructed quantum mechanics based on this generalized probability formalism, where  he called the properties of these generalized probabilities as {\it physical properties} \cite{Hofmann89, Hofmann91}, which provide the full framework of quantum mechanics, including quantum ergodicity \cite{Hofmann89}, and quantum paradoxes \cite{Hofmann91}.

Our main result of this section is to show that the modular value is the average of the dynamic phase factor $e^{-iga}$ over all eigenvalues with the complex conditional probability, which is expressed as 
\begin{align}{\label{mv}}
(\hat{A})_{\rm m} = \sum_a e^{-iga} \text {Pr} (a | \psi,\phi)\;.
\end{align}
To show this, we use the spectral decomposition of an arbitrary function of operator $\hat{A} = \sum_a a |a\rangle\langle a |$, where all eigenvalues $\{|a\rangle\}$ form orthonormal bases. The spectral decomposition of $F(\hat{A})$, where $F$ is an any analytic function, is written as
\begin{align}{\label{f(A) = sum f(a) |a rangle langle a |}}
F(\hat A) = \sum_a F(a) |a\rangle\langle a | \;,
\end{align} 
which is derived by the Taylor (Maclaurin) expansion of function $F(\hat A)$. Choosing $e^{-ig\hat A} $ as the function {$F(\hat{A})$,} Eq. (\ref{f(A) = sum f(a) |a rangle langle a |}) immediately leads to
\begin{align}{\label{SpectralDecompositionForExp(-igA)}}
e^{-ig\hat A} = \sum_a e^{-iga} |a\rangle\langle a|\;.
\end{align} 
Putting this into Eq. (\ref{modular}) and using Eq. (\ref{complex_conditional_prob_1}), we obtain Eq. (\ref{mv}). 

Next, we discuss chain rules. We fix the initial state $|\psi\rangle$ and the final state $|\phi\rangle$, and consider the case that the intermediate state is found to be $|a\rangle$, assuming that there is another intermediate measurement that randomly projects the state onto one of the orthonormal states $|x\rangle$'s. In this case, the chain rule describing the process of taking route $|\psi\rangle\rightarrow|a\rangle\rightarrow|\phi\rangle$ is the summation over all possible $|x\rangle$'s with proper conditional probabilities. Therein, the process of taking route $|\psi\rangle\rightarrow|a\rangle\rightarrow|x\rangle\rightarrow|\phi\rangle$ for each $|x\rangle$ is the product of $\text{Pr}(x|\psi,\phi)$, which is the process of taking $|\psi\rangle\rightarrow|x\rangle\rightarrow|\phi\rangle$ conditioned by the initial $|\psi\rangle$ and the final $|\phi\rangle$, and $\text{Pr}(a|\psi,x)$, which is the process of taking $|\psi\rangle\rightarrow|a\rangle\rightarrow|x\rangle$ conditioned by  $|\psi\rangle$ and $|x\rangle$. Thus  the chain rule  becomes \cite{Hofmann89}
\begin{align}{\label{Bayes-like}}
 \text{Pr}(a|\psi,\phi) &= \sum_x \text{Pr}(a|\psi,x) \text{Pr}(x|\psi,\phi) \;.
\end{align}
Substituting Eq. \eqref{Bayes-like} into Eq. \eqref{Aw=sum_a a Pr(a|psi,phi)}, and using $\text{Pr} (x | \psi,\phi)  = \langle \hat{\Pi}_x \rangle_{\rm w}$ with $ \hat{\Pi}_x \equiv |x\rangle\langle x|$, we obtain 
\begin{align}{\label{weak_rule}}
{_\phi\langle \hat{A}\rangle_\psi^{\rm w}} = \sum_x  {_x\langle A\rangle_\psi^{\rm w}} \cdot {_\phi\langle \hat{\Pi}_x\rangle_\psi^{\rm w}} \;,
\end{align}
where ${_f\langle \cdot\rangle_i^{\rm w}}$ denotes the weak value between pre- and post- selection states $|i\rangle$ and $\langle f|$, respectively. 

In this paper, we obtain a chain rule  for the modular value in a similar way.  In fact, substituting Eq. \eqref{Bayes-like} into Eq. \eqref{mv}, we obtain 
\begin{align}{\label{modular_rule}}
{_\phi(\hat{A})_\psi^{\rm m}} = \sum_x  {_x(\hat{A})_\psi^{\rm m}}\cdot{_\phi\langle \hat{\Pi}_x\rangle_\psi^{\rm w}}\;,
\end{align}
where ${_f(\cdot)_i^{\rm m}}$ denotes the modular value between pre- and post- selection states $|i\rangle$ and $\langle f|$, respectively. We can also generalize the expression if we define
\begin{align}{\label{generalized_modular}}
(\hat{A})_{\rm F} \equiv \dfrac{\langle\phi| F(\hat{A}) |\psi\rangle}{\langle\phi|\psi\rangle}
= \sum_a F(a) \text {Pr} (a | \psi,\phi)\;,
\end{align}
where function $F$ can be any analytic function.  
Substituting Eq. \eqref{Bayes-like} into Eq. \eqref{generalized_modular}, we also obtain the chain rule: 
\begin{align}{\label{generalized_modular_rule}}
{_\phi(\hat{A})_\psi^{\rm F}} = \sum_x  {_x(\hat{A})_\psi^{\rm F}} \cdot{_\phi\langle \hat{\Pi}_x\rangle_\psi^{\rm w}}\;. 
\end{align}
When $F(a) = a$, it leads to the weak value, and when $F(a) = e^{-iga}$, it leads to the modular value. 

\bigskip

{\section{The modulus of modular values}\label{sec_mod}}
We consider the system-pointer interaction Hamiltonian $\hat{H}$, and assume that the interaction is not weak but  can be arbitrarily large. Initially, the quantum system is prepared in $|\psi\rangle$ and the pointer is prepared in $|\xi\rangle$. Following the standard von Neumann treatment, the unitary evolution $\hat U(g)$ for the measurement is assumed to be $\hat U(g) = e^{-ig\hat A^{\rm s}\otimes\hat P^{\rm p}}$, where $\hat A^{\rm s}$ is an operator in the system Hilbert space $\mathcal H^{\rm s}$ and $\hat P^{\rm p} \equiv |\eta \rangle \langle \eta |$ ($|\eta\rangle$ is one of the orthonormal bases of the pointer) is a selected projection operator in the pointer Hilbert space $\mathcal H^{\rm p}$. The role of $|\eta\rangle$ is the same as $|1\rangle$ in the qubit pointer in Sec. \ref{sec_int}, which means that we generalize the qubit pointer to qudit pointer (i.e., the dimension of the pointer Hilbert space is arbitrary but finite.) Now we can calculate the joint transitional probability to find the pointer in $|\mu\rangle$ ($|\mu\rangle$ can be any of the bases, which might or might not be equal to $|\eta\rangle$) and the system in the final state $|\phi\rangle$. We write this conditional joint probability as  ${\rm Pr}_g(\mu,\phi|\xi,\psi)$,  where $\mu$ and $\phi$ are obtained outcomes indicating that the pointer is found in the state $|\mu\rangle$ and the system in $|\phi\rangle$, respectively.   Similarly, $\xi$ and $\psi$ are the observed indicators corresponding to the preparation of the pointer in $|\xi\rangle$ and the system in $|\psi\rangle$, respectively.  The conditional joint probability is calculated to be
\begin{align}{\label{joint_pro}}
{\rm Pr}_g(\mu,\phi|\xi,\psi) = \text{Tr} (\hat M_\mu^\dagger \hat \Pi_f \hat M_\mu \hat \rho_i)\;,
\end{align}
where $\hat \Pi_f = |\phi\rangle\langle\phi|$ and $\hat\rho_i = |\psi\rangle\langle\psi|$, respectively, and the subscript $g$ means that the probability is $g$ dependent. The operator $\hat M_\mu\equiv \langle\mu| e^{-ig\hat A^{\rm s}\otimes\hat P^{\rm p}}|\xi\rangle =  \langle\mu|\xi\rangle e^{-ig\hat A^{\rm s}\delta_{\mu\eta}}$ is known as the {\it Kraus operator}, which is acting on the system Hilbert space $\mathcal H^{\rm s}$. It is easy to check that $\sum_\mu \hat M_\mu^\dagger  \hat M_\mu = \hat I^{\rm s}$. Then, the straightforward calculation of the joint transitional probability \eqref{joint_pro} leads to ${\rm Pr}_g(\mu,\phi|\xi,\psi) = |\langle\mu|\xi\rangle|^2\cdot |\langle\phi|e^{-ig\hat A^{\rm s}\delta_{\mu\eta}}|\psi\rangle|^2 \;.$ 

Now, let us show that this formalism is useful when we try to measure the modular value experimentally. We consider the case that the pointer is a qubit with the initial state $|\xi\rangle = \gamma|0\rangle + \bar\gamma|1\rangle$, where $\gamma$ and $\bar \gamma$ are assumed to be real, satisfying $\gamma^2+\bar\gamma^2=1$. The projection operator $\hat P^{\rm p}$ is chosen to be $|1\rangle\langle 1|$, that is, $|\eta\rangle = |1\rangle$. Then it is straightforward to calculate the joint probability for the cases $\mu = 0$ and 1, that is, for the cases that we find the qubit pointer in $|\mu\rangle = |0\rangle$ and $|1\rangle$, respectively. The results become
\begin{align}{\label{P_g(0)}}
{\rm Pr}_g(0,\phi|\xi,\psi) = \gamma^2|\langle\phi|\psi\rangle|^2 \quad {\rm for} \ \mu = 0 \;,
\end{align}
and
\begin{align}{\label{P_g(1)}}
{\rm Pr}_g(1,\phi|\xi,\psi) = \bar\gamma^2|\langle\phi|e^{-ig\hat A^{\rm s}}|\psi\rangle|^2 \quad {\rm for} \ \mu = 1\;.
\end{align}
Now we introduce the ratio of ${\rm Pr}_g(1,\phi|\xi,\psi)$ to ${\rm Pr}_g(0,\phi|\xi,\psi)$, which we denote by $\chi$.  
Then, using Eqs. (\ref{P_g(0)}) and  (\ref{P_g(1)}), we obtain
\begin{align}{\label{modulus_99}}
\chi \equiv {{\rm Pr}_g(1,\phi|\xi,\psi) \over {\rm Pr}_g(0,\phi|\xi,\psi)} = \left| \dfrac{\bar{\gamma}}{\gamma} \right|^2 \cdot \dfrac{|\langle\phi|e^{-ig\hat A^{\rm s}}|\psi\rangle|^2 }{|\langle\phi|\psi\rangle|^2} \;.
\end{align}
This equation is interesting in the sense that the right-hand side means the relative change in the post-selection probabilities  of finite $g$ case to the $g=0$ case for the system, whereas the definition of $\chi$ is the ratio of the statistical frequency of finding $\mu = 1$ to that of $\mu = 0$ for the pointer. For this reason, we will simply refer to $\chi$ as ^^ ^^ relative change", hereafter. 

Using Eqs. (\ref{modular}) and (\ref{modulus_99}), we obtain the relation between $\chi$ and the modular value, as
\begin{align}{\label{modulus_98}}
\left |(\hat{A}^{\rm s})_{\rm m}\right| = \left| \dfrac{\gamma}{\bar\gamma} \right| \sqrt{\chi}
\;.
\end{align}
The value $\chi$ is easily obtained experimentally because $\chi$ is the average of a series of measurement results for the qubits pointer, and each measurement is done after separating the system and the pointer, and thus, we can use a strong and destructive measurement each time.  
In this way, the modulus of a modular value of the system can be obtained from measurements of the qubits pointer.

It is worth to note that our definition about the relative change $\chi$ is slightly different from that of the Dressel et al. approach, where the authors supposedly defined the relative change as the ratio in between the changing transitional probability induced by the system-pointer interaction and the initial transitional probability without the interaction for weak coupling strength \cite{Dressel86,Dressel91} . With $\gamma =\bar\gamma = 1/\sqrt{2}$, however, our $\chi$ becomes equivalent to the Dressel's one.

Furthermore, in the Dressel et al. works, the authors have shown that for the constant $g$ is small enough, the Taylor series expansion gives the {\it linear dependence} of $\chi$ as $\chi_{\rm w} = 1+ 2g \ \text{Im}\langle \hat{A}\rangle_{\rm w}$ \cite{Dressel86}. Here, we used subscript w for $\chi$ because it is expressed by the weak value. This expression is valid only in the {\it weak interaction regime}, where the higher order terms can be neglected and thus the following requirements are satisfied: (i) the relative change $\chi_{\rm w} $ should be close to one, and (ii) the first order term should be large compared to the sum of all higher order corrections (see Ref. \cite{Dressel86} and Ref. \cite{Duck40} and references therein). The above linear expression of $\chi_{\rm w}$ can be used  for experimental estimation in two ways. One is to estimate the magnitude of $g$ when $g$ is not known but $\chi_{\rm w}$ and the imaginary part of $\langle \hat A \rangle_{\rm w}$ are experimentally obtainable. Another is the indirect estimation of the observable using only the detector statistics \cite{Dressel91}. In both cases, however, we can use $\chi_{\rm w}$ only when $g$ is small enough.

For arbitrary values of $g$, however, our ``relative change" gives the exact relation to the modulus of modular values. Therefore, we can rigorously write $\chi_{\rm m} = \dfrac{\bar\gamma^2}{\gamma^2}\left|(\hat A^s)_{\rm m}\right|^2$, where we use the subscript m for $\chi$ as well since the relative change, in this case, is induced by the modular value rather than the weak value.  So, $\chi_{\rm m}$ enlarges the usage of the ^^ ^^ relative change" much more than $\chi_{\rm w}$. Of course, this expression of $\chi_{\rm m}$ is exact for small $g$ including $g=0$. When we increase $g$, however, $\chi_{\rm m}$ clearly shows its nonlinear dependence on $g$. 

\begin{figure} [t!]
\centering
\includegraphics[width=6.5cm]{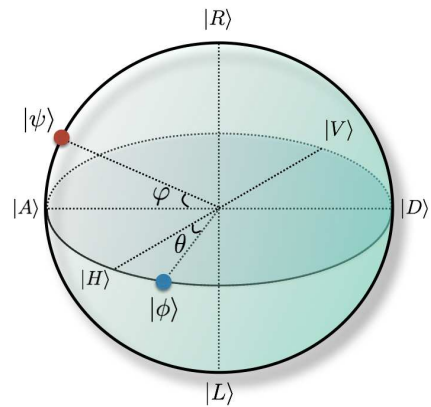}
\caption{
(color online) A schematic drawing of the Bloch (Poncar\'e) sphere exhibiting the qubit states. The lateral angle $\varphi$ of the initial state $|\psi\rangle$ is fixed constant and the final state $|\phi\rangle$ can be rotated by changing $\theta$ from 0 to 2$\pi$. The polarization states of the qubit are represented on the sphere, where $|H\rangle$: horizontal, $|V\rangle$: vertical, $|D\rangle = \frac{1}{\sqrt{2}}(|H\rangle+|V\rangle)$: diagonal, $|A\rangle = \frac{1}{\sqrt{2}}(|H\rangle-|V\rangle)$: anti-diagonal, $|L\rangle = \frac{1}{\sqrt{2}}(|H\rangle+i|V\rangle)$: left circular, $|R\rangle = \frac{1}{\sqrt{2}}(|H\rangle-i|V\rangle)$: right circular.
}
\label{fig1}
\end{figure}

As an illustration, let us consider a polarization qubit ($|0\rangle = |{\rm H}\rangle$: horizontally polarized photon, $|1\rangle = |{\rm V}\rangle$: vertically polarized photon), and assume the initial and final qubit states, following \cite{Dressel86}, to be
\begin{align}{\label{state}}
|\psi\rangle = \dfrac{|{\rm H}\rangle -e^{i\varphi}|{\rm V}\rangle}{\sqrt{2}}; \hspace {0.25cm} |\phi\rangle = \cos \frac{\theta}{2}|{\rm H}\rangle + \sin\dfrac{\theta}{2}|{\rm V}\rangle\;,
\end{align}
where, $\varphi$ is the lateral angle for the prepared state, and $\theta$ is the azimuthal angle for the post-selected state on the Bloch (Poncar\'e) sphere (see Fig. \ref{fig1}). The evolution of the qubit is generated by $\hat{A}^s = \hat{S}$, where $\hat S \equiv |{\rm H}\rangle\langle {\rm H}| - |{\rm V}\rangle\langle {\rm V}|$ is the Stokes polarization operator, whose eigenvectors are $|{\rm H}\rangle$ and $|{\rm V}\rangle$ with the corresponding eigenvalues $1$ and $-1$, respectively. 

The joint transitional probabilities ${\rm Pr}_g(\mu,\phi|\xi,\psi)$ corresponding to the outcomes $\mu = $ H and V are calculated as (in the present example, we use Pr(H) and Pr(V) for short)  
\begin{align}
{\rm Pr}(H) &= \gamma^2(1-\cos\varphi\sin\theta)/2\;,\label{p0}\\
\text{Pr}(V) &=\bar\gamma^2\bigl(1-\cos(2g+\varphi)\sin\theta\bigr)/2\;.\label{pm}
\end{align}
Then, using Eqs.\eqref{modulus_99} and \eqref{modulus_98},  the modulus of the modular value is calculated as

\begin{align}{\label{pmdp}}
\left|(\hat{S})_{\rm m}\right| = \biggl[\dfrac{1-\cos(2g+\varphi)\sin\theta}{1-\cos\varphi\sin\theta}\biggr]^{1/2}\;.
\end{align}

\begin{figure} [t!] 
\centering
\includegraphics[width=8.4cm]{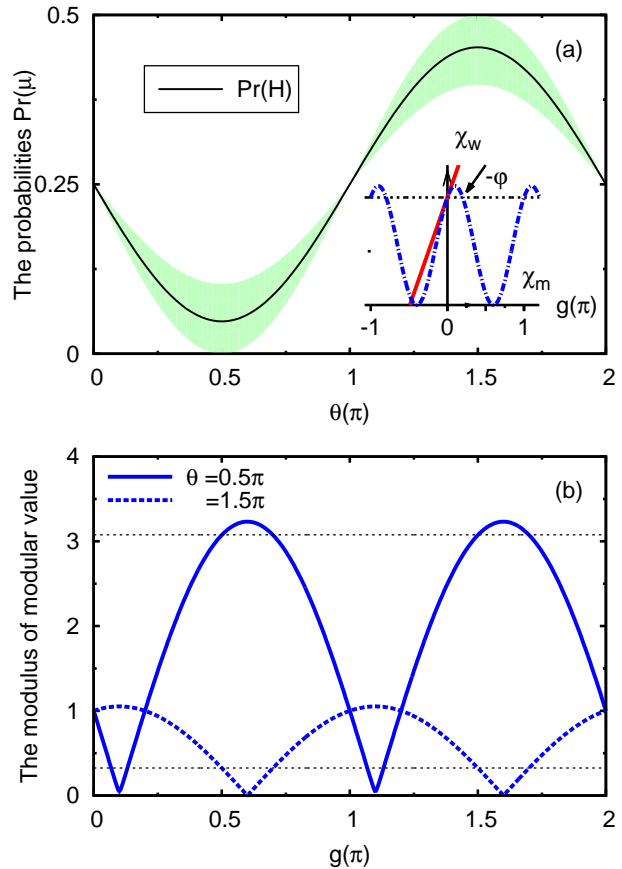}
\caption{
(color online) (a)  Main figure: Plot of the joint transitional probabilities as functions of $\theta$ for various values of $g$ from 0 to 0.25$\pi$. The black solid curve is the Pr(H) and also denotes the Pr(V) with $g = 0$, and $0.2\pi$. The value of $\varphi$ is assumed to be -0.2$\pi$ and $\gamma=\bar\gamma=1/\sqrt{2}$ in all of the plots in (a) and (b). Inset: The $g$ dependence of the relative changes, $\chi_{\rm w}$ (red solid line) and $\chi_{\rm m}$ (blue dash-dotted curve) for $\theta=3\pi/2$. The value 1 is shown by the black dotted horizontal line, which makes it easy to see that $\chi_{\rm m}$ deviates from 1 as $g$ is increased from 0 and then it comes back to 1 again.  The point where this occurs is indicated by the small black arrow, which corresponds to $g = 0.2\pi$ $(= - \varphi)$. 
(b) The $g$ dependence of the modulus of the modular value, $|(S)_{\rm m}|$,  for $\theta = \pi/2$ (blue solid curve) and $\theta=3\pi/2$ (blue dotted curve). The modulus of weak values are constants and shown as the black dotted horizontal lines.}
\label{fig2}
\end{figure}

As a numerical calculation, we fix the value of the lateral angle $\varphi$ of the initial state to be a constant, say, $\varphi = -0.2 \pi$, and assume $\gamma=\bar\gamma=1/\sqrt{2}$. Then we calculate the $\theta$ dependence of the joint transitional probabilities for several values of $g$ and $g$ dependence of the modulus of modular values for several values of $\theta$. Fig. \ref{fig2}(a) shows the $\theta$ dependence of the joint transitional probabilities for $g$ ranging from 0 to 0.25$\pi$. The $\theta$ dependence is sinusoidal both for Pr(H) and Pr(V), but the amplitude depends on $g$ for Pr(V) as is shown by the green area while it does not depend on $g$ for Pr(H) as is shown by the single black curve. 
This black curve also expresses Pr(V) for $g=0$.   By increasing $g$ from 0, Pr(V) gradually deviates  from Pr(H), the deviation becomes maximum at $g=0.1\pi$, and then it comes back to the Pr(H) curve when $g$ reaches to 0.2$\pi$, 
which is shown again by the black curve.

We can see this by plotting $g$ dependence of the relative change $\chi$. The inset of Fig. 2(a) shows the $g$ dependence of the relative changes $\chi_{\rm w}$ and $\chi_{\rm m}$ for $\theta=3\pi/2$. The red line shows the {\it linear} dependence of $\chi_{\rm w}$ to $g$, and therefore, its value takes 1 only when $g$ is 0. On the other hand, the blue dash-dotted curve ($\chi_{\rm m}$) behaves as the {\it sine function} with respects to $g$. The intersections of the black dotted horizontal line (= 1) and $\chi_{\rm m}$ make one easy to see that $\chi_{\rm m}$ deviates from 1 as $g$ is increased from 0 and then comes back to 1 again. This means that the joint transitional probability Pr(V) comes back to the same value as that of the joint transitional probability Pr(H), and its first come-back occurs when $g$ is set to be $g = -\varphi$ ($= 0.2\pi$ in this example), which is indicated by the small black arrow. In addition, whenever the relative change becomes 1, $\chi$ does not change from its original value at all. This no-change points are shown by the intersections between the horizontal line and $\chi_{\rm m}$ (see the inset of Fig. 2a). Interestingly, this no-change-point appears periodically with the period $\pi$. Mathematically, it can be explained by solving the equation Pr(H) - Pr(V) = 0, with respects to $g$. The solution is straightforwardly calculated as follows:
\begin{align}{\label{solu}}
\notag&-\cos(2g+\varphi)\sin\theta+\cos\varphi\sin\theta = 0\\
&\Rightarrow g = 
\left [
	\begin{array}{l}
		k\pi  \\
		-\varphi+k\pi 
	\end{array}
\right.
k \in \mathbb N\;.
\end{align} 

We next analyze the $g$-dependence of the modulus of the modular value $|(S)_{\rm m}|$ shown in Fig. 2(b) for $\theta = \pi/2$ (blue solid curve) and $\theta=3\pi/2$ (blue dashed curve). Here, again, $\varphi$ is chosen to be $\varphi = -0.2\pi$.  The corresponding modulus of the weak values are shown by the black dotted horizontal lines. Obviously, the modulus of the modular value explicitly depends on the coupling constant $g$ and deviates from unity as $g$ is increased from 0. 

The interesting thing is that, as is discussed above, $\chi_{\rm m}$ comes back to 1, and we can find specific values of $g$ $(\ne 0)$ that realizes $\chi_{\rm m} = 1$ again [see example in the inset of Fig. \ref{fig2}(a)]. This 
property of $\chi_{\rm m}$ can be used in the experimental tuning of the value of $g$ to the desired value (such as for preparing the interaction strength in a modular-value measurement). In the following, we show how to do this tuning. Assume that one wants to realize $g=0.3\pi$ by adjusting an uncalibrated experimental set-up. The first thing to do is to prepare the initial state $|\psi\rangle$ of Eq. (\ref{state}) with $\varphi = -0.3\pi$. Starting from no interaction ($g=0$), increase $g$ (not calibrated experimentally yet), and plot $\chi_{\rm m}$ with a certain value of $\theta$. $\theta$ should be chosen so that the deviation of $\chi_{\rm m}$ from 1 is clearly seen. Then, like the inset of Fig. \ref{fig2}(a), $\chi_{\rm m}$ comes back to 1 again. At the very moment when $\chi_{\rm m}=1$ is realized, the set-up is appropriately adjusted to give $g = 0.3\pi$. 

\begin{figure}[t!]
\includegraphics[width=8.6cm]{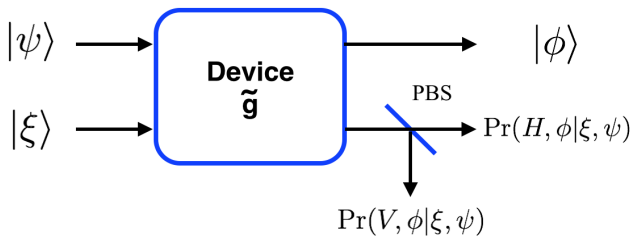}
\caption{
(color online) A schematically set-up of a reference system to determine the value of $g$. At first, the system and qubit pointer states ($|\psi\rangle$, and $|\xi\rangle$) are injected into Device $\widetilde g$, where the interaction is implemented. The system then post-selected onto the final state $|\phi\rangle$, whereas, the qubit pointer output state are passed through the Polarized Beamsplitter (PBS) in order to measure the corresponding probabilities of the Horizontal polarization beam Pr($H,\phi|\xi,\psi$) and the Vertical polarization beam Pr($V,\phi|\xi,\psi$).
}
\label{fig3}
\end{figure} 

To be more concrete, let us consider a reference system as shown in Fig. \ref{fig3}, where Device $\widetilde g$ implements the interaction between a quantum system and a qubit pointer with an uncalibrated but controllable coupling. Prepare a photon in a polarization state $|\psi\rangle$ with the desired value of $\varphi$, and postselect it onto $|\phi\rangle$ as in Eq. \eqref{state}. The initial state of the qubit pointer is $|\xi\rangle = \frac{1}{\sqrt{2}}(|{\rm H}\rangle + |{\rm V}\rangle$). The unitary evolution is $\hat U(g) = e^{-ig\hat S\otimes\hat P}$, where $\hat P\equiv$ $|{\rm V}\rangle\langle {\rm V}|$ denotes the projection operator. In order to determine the value of $g$, one can experimentally change $g$ so that the outcome joint transitional probability $\text{Pr}(H,\phi|\xi,\psi)$ equals to $\text{Pr}(V,\phi|\xi,\psi)$. Then it is guaranteed that the value of $g$ is calibrated to be $-\varphi$.

{\section{The argument of modular values}\label{sec_arg}}
As was discussed by Cormann et al., the argument of a modular value can be measured by the phase in a quantum eraser interference experiment \cite{Cormann}. Here, we analyze the argument of modular values more in detail, and
we show that it is expressed by the Pancharatnam phases.

In 1956, Pancharatnam considered the interference of two optical states, say $|A\rangle$ and $|B\rangle$, that are not orthogonal to each other in polarization \cite{Pancharatnam44}. The Pancharatnam relative phase associated with two states, written as $\delta(A, B)$, is defined by the argument of the inner product of the two states, so, $\delta(A,B) \equiv {\rm arg}[\langle A|B\rangle]$. This is also known as the intrinsic phase \cite{Tamate11}. States $|A\rangle$ and $|B\rangle$ are said to be ``in phase" if the absolute value of the inner product is maximum, i.e. $\delta(A,B)=0$.

This phase does not satisfy the transitive rule, that is, even if both $\delta(A,B)$ and $\delta(B,C)$ are in phase, the relative phase $\delta(A,C)$, in general, is not in phase. This nontransitive property can be seen by considering the three pure states --- ``three" is the smallest nontrivial entity. The Pancharatnam relative phase associated with three states, written as $\Delta( A,B,C)$, is defined by the following equation \cite{Erik359}:
\begin{align}{\label{phase_define}}
\Delta( A,B,C)  \equiv  \text{arg} \Big[\langle A| C\rangle \langle C| B\rangle \langle B| A\rangle \Big]\;.
\end{align}
This phase, which is also known as the geometric phase, is gauge invariant because the local phase factor, which might be independently chosen for each quantum state,  always appears with its complex conjugate due to a couple of bra and ket vectors, and thus all the local phases are canceled. For example in qubit case, this geometric phase is well understood by considering the Bloch (or Poincar\'e) sphere, on which we can draw a geodesic triangle having three vertices, $|A\rangle, |B\rangle$, and $|C\rangle$. It is well established that the geometric phase is expressed by the solid angle $\Omega$ of the geodesic triangle as $\Delta(A,B,C) = -\Omega / 2$. \cite{Pancharatnam44,Ramaseshan55,Berry34}.

As  was discussed in Ref. \cite{Cormann}, the phase in the modular value is an intrinsic property of the quantum system in the sense that the evolution $e^{-ig\hat{A}}$ in the modular value solely depends on the system evolution but not on the measurement apparatus or environments. In this paper, we show that the argument of a modular value is expressed by the summation of an intrinsic phase and a geometric phase, as is shown in the following.

Let us consider the following state-evolution process: an initial state, i.e., pre-selected $|\psi\rangle$, evolves 
under the evolution operator $e^{-ig\hat A}$, then we project the resultant state $|\psi(g)\rangle \equiv e^{-ig\hat A} |\psi\rangle$ onto the post-selection state $|\phi\rangle$ that we selected. The final state (after post-selection) is given by $|\psi_\phi(g)\rangle\equiv|\phi\rangle \langle\phi|e^{-ig\hat A}|\psi\rangle$ (not normalized). The phase difference between this final state and the initial state is calculated to be
\begin{align}{\label{phase_1_rew}}
\notag{\rm arg}[\langle\psi|\psi_\phi(g)\rangle] &= {\rm arg}[\langle\psi|\phi\rangle \langle\phi|e^{-ig\hat A}|\psi\rangle] \\
&={\rm arg}[(A)_{\rm m}].
\end{align}
This is one interpretation of the meaning of the (argument of the) modular value.  Next, we relate this to the geometric phase among $|\psi\rangle, |\psi(g) \rangle$ and $|\phi\rangle$, which is calculated from Eq. \eqref{phase_define} by replacing $|A\rangle, |B\rangle$ and $|C\rangle$ by
$|A\rangle=|\psi\rangle$, $|B\rangle= |\psi(g)\rangle$ and $|C\rangle=|\phi\rangle$ \cite{Erik359, Berry34,Mukunda228}, as 
\begin{align}{\label{phase1}}
\Delta (\psi,\psi(g), \phi) = \text{arg} \Big[\langle \psi|\phi\rangle \langle \phi|\psi(g)\rangle \langle \psi(g)|\psi\rangle\Big]\;.
\end{align}
The meaning of this equation is that this phase shift is induced by the closed-loop projection, $|\psi\rangle \rightarrow |\psi(g)\rangle \rightarrow |\phi\rangle \rightarrow |\psi\rangle$. Using this, the argument of the modular value is calculated to be
\begin{align}{\label{phase_1}}
\notag{\rm arg}[(A)_{\rm m}] &= {\rm arg}\Bigl[\dfrac{\langle\phi|e^{-ig\hat A}|\psi\rangle}{\langle\phi|\psi\rangle}\Bigr]\\
\notag&= {\rm arg}\Bigl[\dfrac{\langle \psi|\phi\rangle \langle\phi|\psi(g)\rangle\langle\psi(g)|\psi\rangle}{|\langle\psi|\phi\rangle|^2\langle\psi(g)|\psi\rangle}\Bigr]\\
\notag&= {\rm arg}\Big[\langle \psi|\phi\rangle \langle \phi|\psi(g)\rangle \langle \psi(g)|\psi\rangle\Big]-\text{arg}[\langle\psi(g)|\psi\rangle]\\
&= \Delta (\psi,\psi(g), \phi) + \delta(\psi,\psi(g))\;.
\end{align}
\begin{figure} [t!] 
\centering
\includegraphics[width=8.6cm]{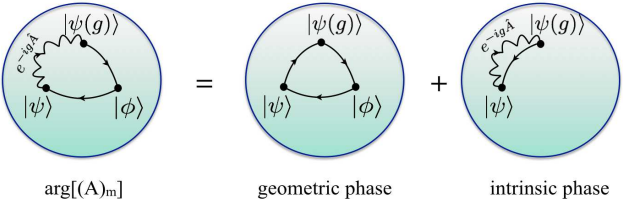}
\caption{
(color online) The argument of a modular value is the total phase shift of a state-evolution process that starting from an initial state, evolve onto an intermediate state, then project onto a final state and finally project back onto the initial state. This total phase shift corresponds to the summation of the geometric phase and the intrinsic phase.
}
\label{fig4}
\end{figure}
Although $|\psi(g)\rangle$ is the result of the evolution induced by $e^{-ig\hat A}$, 
$\Delta$ does not carry any phase shift by this evolution since the bra and ket vectors cancel out this phase shift. 
So, only the pre-post-projections yield the geometric phase $\Delta$ but the evolution does not.  
The evolution phase shift is solely carried by $\delta$.  
In both cases ($\Delta$ and $\delta$), one needs to project the final state onto the initial state to compare the phase shift directly. We illustrate this situation in Fig. \ref{fig4}. Interestingly, as we show here, by picking up a point in the evolution path, the argument of modular value now becomes the summation of the geometric phase $\Delta$ of a geodesic triangle having three vertices $|\psi\rangle$, $|\psi(g)\rangle$, and $|\phi\rangle$, and the intrinsic phase $\delta$ between the initial state and the intermediate state $|\psi(g)\rangle$. This is the main result of this section. 

Particularly, when the coupling constant $g$ is sufficiently small, we can take the first order of the Taylor series expansion of the exponential term and obtain 
\begin{align}{\label{phase_1_gsmall}}
\notag{\rm arg}[(A)_{\rm m}] \big |_{g\to 0}  &\approx {\rm arg}\Bigl[\dfrac{\langle\phi|(\hat I -ig\hat A)|\psi\rangle}{\langle\phi|\psi\rangle}\Bigr]\\
\notag&= {\rm arg}\Bigl[1-ig\langle A\rangle_{\rm w}\Bigr]\\
\notag&\approx {\rm arg}\Big[e^{-ig\langle A\rangle_{\rm w}}\Big]\\
&= -g{\rm Re}\langle A\rangle_{\rm w}\;.
\end{align}
We emphasize that for small $g$, the argument of the modular value does not reduce to the argument of the weak value, instead, it is proportional to the real part of the weak value.

To show the difference and advantage of the argument of modular values, we will compare with the more familiar weak-value case. We consider the weak value of the projection operator $\hat A = |a\rangle\langle a|$, then the argument of the weak value reduces to the geometric phase as:
\begin{align}{\label{phase_weak}}
{\rm arg}[\langle A\rangle_w]=\Delta (\psi,a, \phi) = \text{arg} \big[\langle \psi|\phi\rangle \langle \phi|a\rangle \langle a|\psi\rangle\big],
\end{align}
as is written in \cite{Erik359} and can be measured by the shift in position and momentum of the pointer. This means that, when the process is such a one that the initial state $|\psi\rangle$ is projected onto an intermediate state $|a\rangle$, successively projected onto $|\phi\rangle$, and finally projected onto the initial state $|\psi\rangle$, then the total phase shift is expressed by the geometric phase $\Delta(\psi,a,\phi)$. This process is different from the state-evolution process above.

\bigskip

{\section{Conclusions}\label{sec_con}}
The main purpose of this work is to examine the property of quantum modular values as complex numbers by using the spectral decomposition and the polar decomposition. We interpreted the complex modular value in connection with the complex conditional probability and expressed the modulus component of the quantum modular value by the relative change in the qubit pointer post-selection probabilities, and also the argument component of the quantum modular value by the relation to the Pancharatnam phases.

First, we have considered the chain rule of conditional probabilities in the situation that the system is initially prepared as $|\psi\rangle$, evolves into ${\rm e}^{-{\rm i}g\hat{A}}|\psi\rangle$ ($\hat{A} \equiv \sum_i^N a_i |a_i\rangle\langle a_i|$), is then weakly measured on $\hat{A}$, is then weakly measured on $\hat{X} \equiv \sum_j^N x_j|x_j\rangle\langle x_j|$, and  is finally projected on $|\phi\rangle$. The result is Eq. (\ref{modular_rule}), which means that the modular value is the sum of the products {over $j$}, each of which is the product of the modular value of $\hat{A}$ sandwiched by $\langle\psi|$ and $|x_j\rangle$ and the weak value of $|x_j\rangle\langle x_j|$ sandwiched by $\langle\psi|$ and $|\phi\rangle$. This is a generalization of the weak-value chain rule, Eq. (\ref{weak_rule}). The most generalized expression is also given as Eq. (\ref{generalized_modular_rule}), where $F(a) = a$ gives the weak-value chain rule, and $F(a) = {\rm e}^{-{\rm i}ga}$ gives the modular-value chain rule. In this sense, the modular value, like the weak value, plays the role of complex conditional probability. 

Next, we have investigated the modulus of modular values, and obtained an expression that describes $|(A)_{\rm m}|$ being proportional to the square root of the ratio of the joint transitional probabilities of the qubit pointer, which we refer to as the relative change in the qubit pointer post-selection probabilities, $\chi$, as is described in Sec. \ref{sec_mod}.  Focus on this relation, we can see that, for small $g$ case (i.e., small evolution case), $\chi$ deviates from 1 linearly as we increase $g$ from 0. We have shown that, however, $\chi$ shows the nonlinear behavior to $g$, and in some particular cases, such as $\gamma=\bar\gamma=1/\sqrt{2}$, $\chi$ comes back to 1 at some points of the value $g=g'$. Using this value $g'$,  especially for the qubit systems, we can experimentally calibrate the coupling parameter $g$ by watching the shift of the relative change in the joint transitional probabilities.

Finally, we have obtained a relation that connects the argument of modular values to the summation of the geometric phase of a closed triangle on the Bloch (Poincar\'e) sphere spanned by $|\psi\rangle$, ${\rm e}^{-{\rm i}g\hat{A}}|\psi\rangle$, and $|\phi\rangle$, and the intrinsic phase spanned by $|\psi\rangle$, and ${\rm e}^{-{\rm i}g\hat{A}}|\psi\rangle$, as is described in Sec. \ref{sec_arg}.

Since the modulus of the modular value is related to the relative change in the joint transitional probabilities (Sec. \ref{sec_mod}), and the argument of the modular value is related to the Pancharatnam intrinsic phase and the geometric phase (Sec. \ref{sec_arg}), both the modulus and the argument components of the modular value are connected to the experimentally obtainable quantities.

\bigskip 

\begin{acknowledgments}
This work was supported by JSPS Grant-in-Aid for Scientific Research(A) JP16H02214.
\end{acknowledgments}

\bibliography{basename of .bib file}

\end{document}